\documentclass[11pt]{article}
\usepackage{graphicx}
\usepackage{amssymb,amsmath,amsfonts}

\def\Re{{\cal R \mskip-4mu \lower.1ex \hbox{\it e}\,}}
\def\Im{{\cal I \mskip-5mu \lower.1ex \hbox{\it m}\,}}
\def\ie{{\it i.e.}}
\def\eg{{\it e.g.}}

\def\etal{{\it et al.}}

\def\sub#1{_{\lower.25ex\hbox{$\scriptstyle#1$}}}
\def\tev{\,{\ifmmode\mathrm {TeV}\else TeV\fi}}
\def\gev{\,{\ifmmode\mathrm {GeV}\else GeV\fi}}
\def\mev{\,{\ifmmode\mathrm {MeV}\else MeV\fi}}
\def\mpl{\ifmmode M_{pl}\else $M_{pl}$\fi}
\def\mpl{\ifmmode \overline M_{Pl}\else $\bar M_{Pl}$\fi}
\def\to{\rightarrow}

\def\subw{_{\rm w}}
\def\mh{\ifmmode m\sbl H \else $m\sbl H$\fi}
\def\mch{\ifmmode m_{H^\pm} \else $m_{H^\pm}$\fi}
\def\mt{\ifmmode m_t\else $m_t$\fi}
\def\mc{\ifmmode m_c\else $m_c$\fi}
\def\mz{\ifmmode M_Z\else $M_Z$\fi}
\def\mw{\ifmmode M_W\else $M_W$\fi}
\def\mws{\ifmmode M_W^2 \else $M_W^2$\fi}
\def\mhs{\ifmmode m_H^2 \else $m_H^2$\fi}   
\def\mzs{\ifmmode M_Z^2 \else $M_Z^2$\fi}
\def\mts{\ifmmode m_t^2 \else $m_t^2$\fi}
\def\mcs{\ifmmode m_c^2 \else $m_c^2$\fi}
\def\mchs{\ifmmode m_{H^\pm}^2 \else $m_{H^\pm}^2$\fi}
\def\ztwo{\ifmmode Z_2\else $Z_2$\fi}
\def\zone{\ifmmode Z_1\else $Z_1$\fi}
\def\mtwo{\ifmmode M_2\else $M_2$\fi}
\def\mone{\ifmmode M_1\else $M_1$\fi}
\def\tb{\ifmmode \tan\beta \else $\tan\beta$\fi}
\def\xw{\ifmmode x\subw\else $x\subw$\fi}
\def\ch{\ifmmode H^\pm \else $H^\pm$\fi}
\def\lum{\ifmmode {\cal L}\else ${\cal L}$\fi}
\def\inpb{\,{\ifmmode {\mathrm {pb}}^{-1}\else ${\mathrm {pb}}^{-1}$\fi}}
\def\infb{\,{\ifmmode {\mathrm {fb}}^{-1}\else ${\mathrm {fb}}^{-1}$\fi}}
\def\epem{\ifmmode e^+e^-\else $e^+e^-$\fi}
\def\ppb{\ifmmode \bar pp\else $\bar pp$\fi}
\def\bsg{\ifmmode B\to X_s\gamma\else $B\to X_s\gamma$\fi}
\def\bsll{\ifmmode B\to X_s\ell^+\ell^-\else $B\to X_s\ell^+\ell^-$\fi}
\def\bstt{\ifmmode B\to X_s\tau^+\tau^-\else $B\to X_s\tau^+\tau^-$\fi}
\def\lamt{\ifmmode \tilde\lambda\else $\tilde\lambda$\fi}
\def\shat{\ifmmode \hat s\else $\hat s$\fi}
\def\that{\ifmmode \hat t\else $\hat t$\fi}
\def\uhat{\ifmmode \hat u\else $\hat u$\fi}

\newskip\zatskip \zatskip=0pt plus0pt minus0pt
\def\matth{\mathsurround=0pt}
\def\lsim{\mathrel{\mathpalette\atversim<}}

\def\atversim#1#2{\lower0.7ex\vbox{\baselineskip\zatskip\lineskip\zatskip
  \lineskiplimit 0pt\ialign{$\matth#1\hfil##\hfil$\crcr#2\crcr\sim\crcr}}}

\def\grtsim{\,\,\rlap{\raise 3pt\hbox{$>$}}{\lower 3pt\hbox{$\sim$}}\,\,}
\def\lsim{\,\,\rlap{\raise 3pt\hbox{$<$}}{\lower 3pt\hbox{$\sim$}}\,\,}

\renewcommand{\thefootnote}{\fnsymbol{footnote}}

\begin{document} \begin{titlepage}
\rightline{\vbox{\halign{&#\hfil\cr
&SLAC-PUB-13572\cr
}}}
\begin{center}
\thispagestyle{empty} \flushbottom { {
\Large\bf Indirect Searches for $Z'$-like Resonances at the LHC  
\footnote{Work supported in part
by the Department of Energy, Contract DE-AC02-76SF00515}
\footnote{e-mail:
rizzo@slac.stanford.edu}}}
\medskip
\end{center}

\centerline{Thomas G. Rizzo}
\vspace{8pt} 
\centerline{\it SLAC National Accelerator Laboratory, 2575 Sand Hill Rd., Menlo Park, CA, 94025}

\vspace*{0.3cm}

\begin{abstract}
We explore the possibility of indirectly observing the effects of $Z'$-like particles with electroweak strength couplings in the Drell-Yan channel at the LHC 
with masses above the resonance direct search reach. We find that, mostly due to statistical limitations, this is very unlikely in almost all classes of models 
independently of the spin of the resonance. Not unexpectedly, the one possible exception to this general result is the case of degenerate Kaluza-Klein (KK) excitations 
of the photon and $Z$ that occur in some extra-dimensional models. In this special case, the strong destructive interference with the Standard Model (SM) exchanges 
below the resonance mass leads to a well-known significant suppression of the cross section and thus increased sensitivity to this particular new physics scenario.   
\end{abstract}


\renewcommand{\thefootnote}{\arabic{footnote}} \end{titlepage} 

%
%
%

\section{Introduction and Background}

The LHC is turning on later this year and the exploration of the TeV scale will commence soon thereafter. One of the cleanest signals for new physics that 
may be discovered early on is the existence of a $Z'$-like resonance in the Drell-Yan channel. Such resonance structures, with a wide variety of possible 
identities, are predicted to exist in a large number of extensions to the Standard Model(SM){\cite {rev}}. Presently, for states with typical electroweak strength  
couplings to SM fields, 
current constraints from the Tevatron restrict their masses to lie above $\sim 0.8-1$ TeV{\cite {tev}}. Once large integrated luminosities of order 
$\sim 100-300$ fb$^{-1}$ are obtained at the LHC, these new resonances should be visible for masses as large as $\sim 4-6$ TeV depending on the specific details of 
the model. The only way to directly access states with larger masses at the LHC is to significantly increase the integrated luminosity into the few ab$^{-1}$ range 
as might be expected at the SLHC upgrade. Eventually, however, these direct searches will run out of steam resulting in a limited mass reach. Is there any way to 
see beyond this direct search reach at the LHC? 

It is well-known that future $e^+e^-$ colliders operating in the TeV energy range can indirectly probe for new $Z'$ resonances by exploring contact interaction-like 
deviations from the cross sections and asymmetries predicted by the SM{\cite {ilc}} in various final state channels. For the typical luminosities that are often 
discussed for these colliders, $\sim 0.5-1$ ab$^{-1}$, and making use of electron and positron beam polarizations, the  masses of $Z'$-like states as large as 
$\sim 10-20 \sqrt s$ may be explored in this way. If these new resonance states are not too heavy it may also be possible to extract the corresponding resonance 
mass itself ({\it without} prior knowledge from the LHC) using 
$e^+e^-$ data collected at several different center of mass energies{\cite {old}}. Unfortunately, it seems more than likely that such colliders will not become 
available until at least the end of the 2020's. In the meantime we might wonder if the mass reach of the (S)LHC itself could be extended in any similar way. 
In this paper we will 
explore the possibility of indirectly observing the effects of such $Z'$-like states which are too massive to be seen directly as resonances at the the LHC, somewhat 
in analogy with what is done at $e^+e^-$ colliders, employing data taken over a range of dilepton masses. Clearly such an analysis is only applicable to resonances 
with couplings to the SM fermions which are of electroweak strength or larger. We note in passing that lighter Drell-Yan resonance states with very small couplings, 
perhaps missed by the Tevatron, will be difficult if not impossible to observe indirectly at the LHC and so will not be a subject for discussion here. 

To some extent, if we are exploring dilepton invariant masses {\it far} below the new resonance, one might imagine that much of this indirect search information 
could be 
extracted directly from the more conventional contact-interaction studies at the LHC{\cite {example}} and this is indeed the case to some extent. However, in the 
more interesting (and, perhaps, more likely) case that we will consider here, the mass of the resonance is not too far above the $5\sigma$ discovery reach of the LHC  
so that non-interference terms as well as the finite resonance width effects also become important thus warranting a separate study. In addition to this, we note that 
traditional contact-interaction analyses are usually restricted to (combinations of) the familiar chiral, dimension-6, four-fermion operators introduced long 
ago{\cite {mp}} and do not allow for the possibility of other spin combinations that we must in all generality consider below.

\section{Analysis}

Given the rather wide selection of possible models with Drell-Yan resonances having electroweak strength couplings to the SM fermions, we will employ four sample  
benchmark examples in the analysis that follows: ($i$) a conventional $Z'$ with electroweak couplings. Here we will employ the specific $Z'$ of the Sequential SM (SSM) 
which is a standard candle for experimenters; in this case the $Z'$ is simply the SM $Z$ but with a significantly larger mass. Other, more realistic, $Z'$ states 
with couplings of comparable magnitude will yield very similar results since this example captures the essential properties of this class of spin-1 models. ($ii$)  
A Randall-Sundrum(RS){\cite {rs}} graviton with the SM fields localized to the TeV brane; to be specific we will assume that $k/\mpl=0.04$ as a typical value for 
purposes of demonstration. Alternative values of this parameter will, of course, modify the signal rate we would obtain. 
This model allows us to examine the case of a spin-2 resonance which corresponds to a dimension-8 four-fermion interaction far below the (first) graviton mass. Of 
course the complete graviton Kaluza-Klein(KK) spectrum must be included in such an analysis; here for practical reasons we will sum over the first dozen KK states 
which provide the dominant contributions to the scattering amplitude in the dilepton mass range of interest at and below the first resonance peak. Similarly, 
($iii$) we consider an $R-$parity violating sneutrino, $\tilde \nu$, which we choose for simplicity to have identical electromagnetic strength lepton number violating 
couplings{\cite {snu}} to both quarks and leptons as a typical example of a spin-0 resonance. Other choices would lead to modifications in the predicted rates below. 
It is important to remember that in this spin-0 case there is no 
interference with the usual SM exchanges. Similarly, in the KK graviton spin-2 case above, the interference with the SM exchanges also vanishes when one integrates 
over the full angular range. Finally, we consider ($iv$) the case of simultaneous, TeV-scale, Kaluza-Klein excitations of both the photon and $Z$ with the SM 
fermions being 4-D{\cite {big}} fields localized at the origin of the fifth dimension. In this case each of the resonances have the same couplings as do their SM 
counterparts except for a rescaling by a factor of $\sqrt 2$ due to the normalization of their bulk wavefunctions. This is a very special situation due to the 
well-known, but somewhat unusual, destructive interference that takes place between the SM ($\gamma,Z$) and full set of KK exchange amplitudes{\cite {study}} 
which causes a drastic reduction in the cross section with dilepton invariant masses near $\sim 0.55$ that of the first KK resonances. {\footnote {This is simply the 
result of being in the mass range above the SM $Z$ yet below the KK mass so that the SM and KK amplitudes have nearly identical structure but opposite signs, the KKs 
being just heavy copies of the SM states. This 
destructive interference is difficult to duplicate in other model scenarios even if multiple exchanges were present since a very fine tuning of couplings would be 
required.}}  As in the RS graviton KK case, in practical calculations we will 
sum over the first dozen resonance states whose contributions will dominate in the dilepton mass region of interest. We note that these are only benchmark models; in 
some cases, \eg, that of the R-parity violating sneutrino, the overall coupling strength has been chosen to make the resonance `interesting' in the mass range we 
are studying.

To be specific in the analysis that follows, we will only consider the LHC with $\sqrt s=14$ TeV and with an integrated luminosity of 300 fb$^{-1}$. However, the 
results that we obtain can be straightforwardly translated to other effective values of $\sqrt s$ and integrated luminosity. Furthermore, the CTEQ6.6 
PDF's{\cite {cteq}} will be employed throughout the analysis below in obtaining all of our numerical results as will the usual mass-scale-dependent 
NLO/NNLO K-factors{\cite {kfactors}} whose typical value is $\simeq 1.25$ in the mass range of interest. The first step in our 
analysis is to approximately determine the usual $5\sigma$ direct discovery reach for the resonances in the benchmark set. This can be easily done by examining  
and extending the results from the detector collaborations themselves{\cite {atlas,cms}}. Once these values are determined, we will consider the corresponding 
resonances with somewhat larger masses. 

\begin{figure}[htbp]
\centerline{
\includegraphics[width=8.5cm,angle=90]{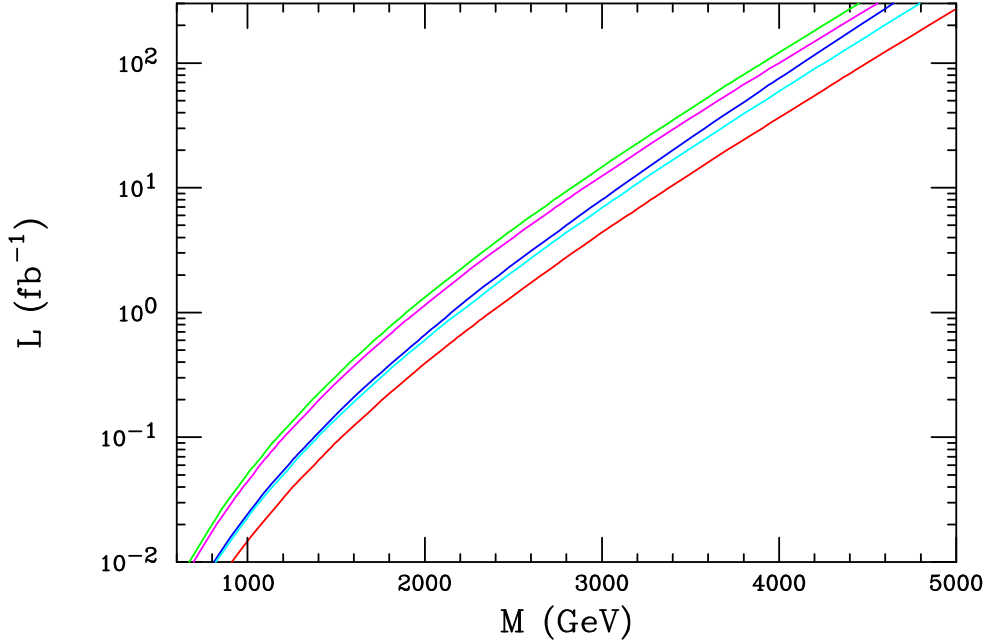}}
\caption{$5\sigma$ resonance discovery reach for conventional $Z'$ states at the 14 TeV LHC as a function of the integrated luminosity in the dilepton channel. 
The red(green, blue, magenta, cyan) curves correspond to the $Z'_{SSM}$($Z'_\psi$, $Z'_\chi$, $Z'_\eta$, $Z'_{LRM}$) model cases, respectively.}
\label{fig0}
\end{figure}

An example of this reach for conventional $Z'$ models is shown in Fig.~\ref{fig0} as a function of the 14 TeV LHC 
integrated luminosity.  Here we have assumed an approximate $100\%$ lepton ID efficiency in this and in all figures to follow. Note that 
a detector smearing of $\sim 1\%$, similar to that of ATLAS for electrons and CMS for muons in this mass range, has been included and, although no $p_T$ cut 
is applied, the final state leptons are required to be central, \ie, with a pair rapidity  
$|y| \leq 2.5$ in all cases. We will employ this cut in all cases below except where as 
otherwise noted; these results are found to match very closely those quoted by both experimental 
collaborations. Here we see that in the SSM $Z'$ model case the LHC discovery reach is $\sim 5$ TeV assuming a luminosity of 300 fb$^{-1}$. In a similar fashion,  
updating the estimates from our various earlier analyses, we can ascertain the $5\sigma$ discovery reaches for the other benchmark models described above. 
Conservatively, we find values of $\sim 4$ TeV for the lightest RS graviton{\cite {dhr}}, $\sim 5.5$ TeV{\cite {taj}} for the $R$-parity violating $\tilde \nu$ and 
$\sim 5.5-6$ TeV{\cite {me}} for the first photon-$Z$ gauge KK states, respectively. 

What observables are available to search for new resonances indirectly at the LHC? There are essentially two: the dilepton invariant mass distribution, $dN/dM$, 
itself and the forward-backward asymmetry, $A_{FB}$, which is determined by the dilepton angular distribution. Here we focus on $A_{FB}$ instead of the angular 
distribution itself since, as we will see, there are already significant statistical issues that we must face. Conventionally, the scattering angle for the 
$q\bar q \to l^+l^-$ process is defined in the center-of-mass frame (where the leptons are back-to-back) to be that 
between the incoming quark and the outgoing negatively charged lepton. At the LHC this quantity cannot be directly measured since we do not know from which 
proton the quark came{\cite {rev}}. However, this problem can be somewhat circumvented as quarks generally carry more momentum in the proton than do anti-quarks 
and thus the direction of motion, \ie, boost, of the dilepton system in the lab frame is more likely to be that of the incoming quark. Since this approach can 
sometimes incorrectly 
assign the quark direction, the resulting raw asymmetry is diluted and this must be corrected for in the analysis via Monte Carlo{\cite {corr}}. (We will assume that 
this can be done correctly and with high efficiency in all cases in our discussion below). Correspondingly, this approach requires us to place a lower bound on the 
rapidity of the outgoing leptons in high mass events in order to define forward and backward hemispheres to obtain $A_{FB}$. Here one traditionally 
requires{\cite {rev}} that the lepton pair rapidity 
$|y|\geq 0.8$ which removes very central events in the detector, but also results in reduced statistics thus increasing the 
corresponding error on the value of $A_{FB}$. 

This is actually a more serious issue than 
one may initially imagine as a further problem with $A_{FB}$ develops for the case of the very high mass dilepton pairs that we are considering: 
the collider kinematics leads to an upper limit on the lepton pair rapidity (in addition to the usual experimental cut of $|y|\leq 2.5$) given 
by the expression found in  Ref.{\cite {ehlq}}:       
\begin{equation}
|y| \leq log ({\sqrt s}/M)\,. 
\end{equation}
Note that in the Drell-Yan process the fractional momenta of the initial state partons, which must satisfy $sx_1x_2=M^2$, are given{\cite {ehlq}} by 
$x_{1,2}=(M/\sqrt s)e^{\pm y}$; requiring that $0\leq x_{1,2} \leq 1$ immediately leads to the constraint equation above. Thus, ever larger values of $M$ for 
fixed $\sqrt s$ force the final state leptons to be more and more central.  
Since here $\sqrt s=14$ TeV, taking $M=3(4,5,6)$ TeV as will be relevant below, leads to the constraint $|y|\leq 1.54(1.25,1.03,0.85)$, respectively. Since we 
are simultaneously requiring that $|y|\geq 0.8$ to obtain a reasonably undiluted $A_{FB}$ in the first place it is clear that the available phase space is 
very rapidly shrinking at high masses leading to rather poor available statistics. As we will see below, this implies that $A_{FB}$ will not play a major role in 
indirect high mass resonance searches at the LHC even if we allow for a somewhat softer lower cut on the lepton rapidity.  

\begin{figure}[htbp]
\centerline{
\includegraphics[width=8.5cm,angle=90]{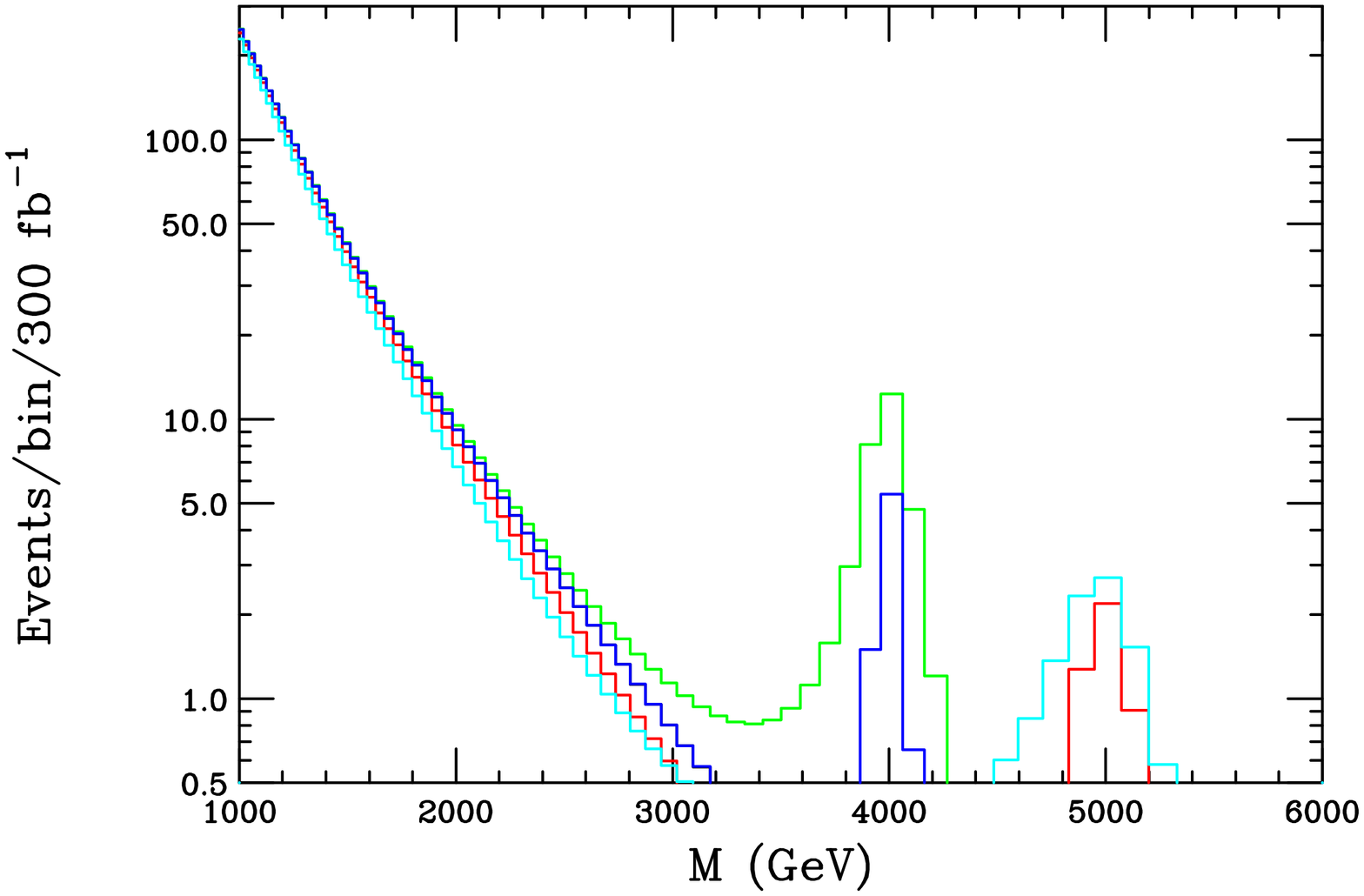}}
\vspace*{0.4cm}
\centerline{
\includegraphics[width=8.5cm,angle=90]{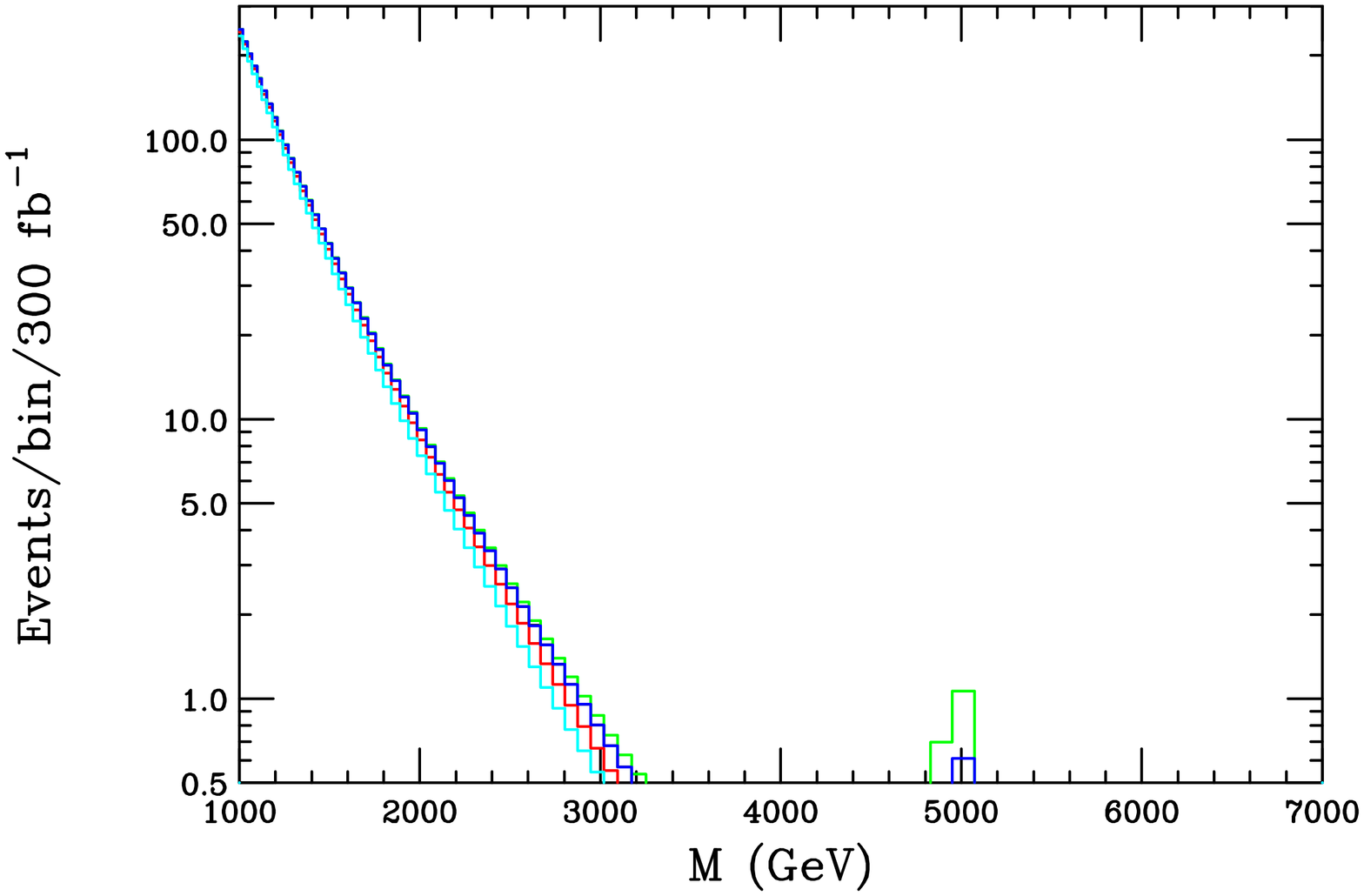}}
\caption{Top panel: Idealized dilepton invariant mass distributions, neglecting the effects of statistical fluctuations(\ie, finite statistics accounting only for the 
overall rate and not the $\sim \sqrt N$ uncertainty effects in any given bin), 
for the SM(black), the SSM with a 5 TeV $Z'$(red), a photon/$Z$ gauge KK state with a mass of 5 TeV(cyan), an 
RS graviton with a 4 TeV mass(blue) and a 4 TeV $R$-parity violating sneutrino(green). Bottom panel: Same as above but now with all the resonance masses increased by 
1 TeV. A detector smearing of $\sim 1\%$, similar to that of ATLAS for electrons and CMS for muons in this mass range, has been included and the final state leptons are 
required to be central, \ie, $|y| \leq 2.5$ in all cases as discussed in the text. Unless otherwise stated our bin width is $1\%$ of the central mass value 
for each bin and NLO/NNLO K-factors are included.}
\label{fig1}
\end{figure}

Let us begin our analysis by considering the invariant mass distribution for Drell-Yan produced dileptons, $dN/dM$. A straightforward calculation including K-factors, 
efficiencies, experimental cuts and approximate detector smearing as discussed above but {\it neglecting} the effect of bin-to-bin $\sim \sqrt N$ statistical fluctuations 
(associated with the limited number of events in each bin and causing the apparent `up and down' bin to bin changes in the plot) leads to the results 
shown in Fig.~\ref{fig1}. Here we see several things: first, this figure verifies to some extent our previous estimates of the $5\sigma$ search reaches for the 
set of benchmark models that we are considering. Second, for the cases where the resonances are near or above the direct search reach, we can easily observe in 
some cases what 
appear to be deviations from the SM predictions for the $dN/dM$ distribution in the dilepton mass range from $\sim 2.5-4$ TeV, somewhat below the resonance masses 
themselves. If these deviations were statistically significant (neglecting for the moment any discussion of systematic errors arising from, \eg, PDF uncertainties and 
electroweak corrections) by fitting the shape of this deviation to a model-dependent excitation curve we would obtain a smoking gun for the existence of a new resonance 
as well as a possible estimate of its mass. It is important to remember that in obtaining these results we have assumed the coupling parameters as described above. The 
fact that, \eg, the sneutrino is more easily visible than is the RS graviton is merely due to the choices of couplings we have made above for the assumed values 
of the resonance masses under study for the different models. What we have shown by this analysis is that, in the absence of statistical fluctuations, some resonances 
will produce visible deviations in the dilepton mass distribution from the SM expectations in the region below the resonance mass. As we will now show, these 
fluctuations lead to a `washing-out' of this effect rendering indirect observation essentially impossible.  

\begin{figure}[htbp]
\centerline{
\includegraphics[width=8.5cm,angle=90]{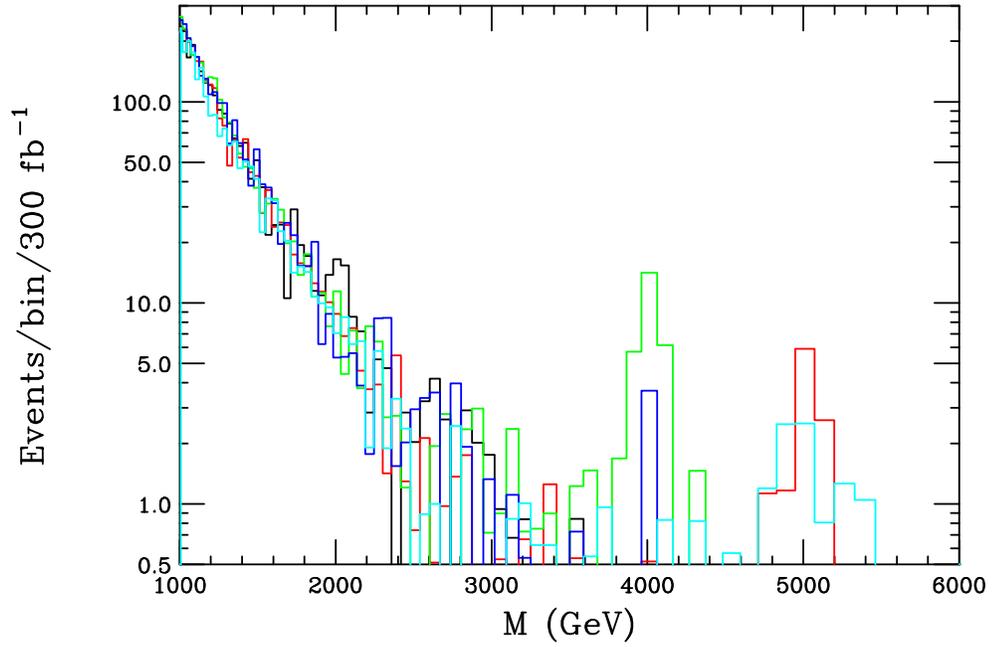}}
\vspace*{0.4cm}
\centerline{
\includegraphics[width=8.5cm,angle=90]{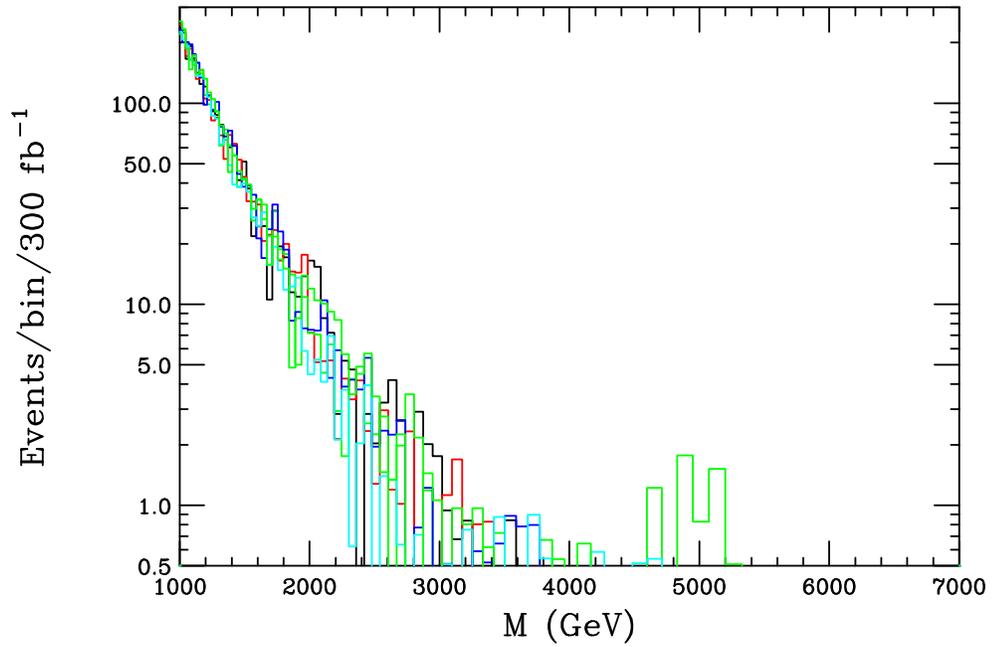}}
\caption{Same as the previous figure but now showing the effect of the bin-to-bin $\sim \sqrt N$ statistical fluctuations 
in the number of events due to finite luminosity.}
\label{fig2}
\end{figure}

\begin{figure}[htbp]
\centerline{
\includegraphics[width=8.5cm,angle=90]{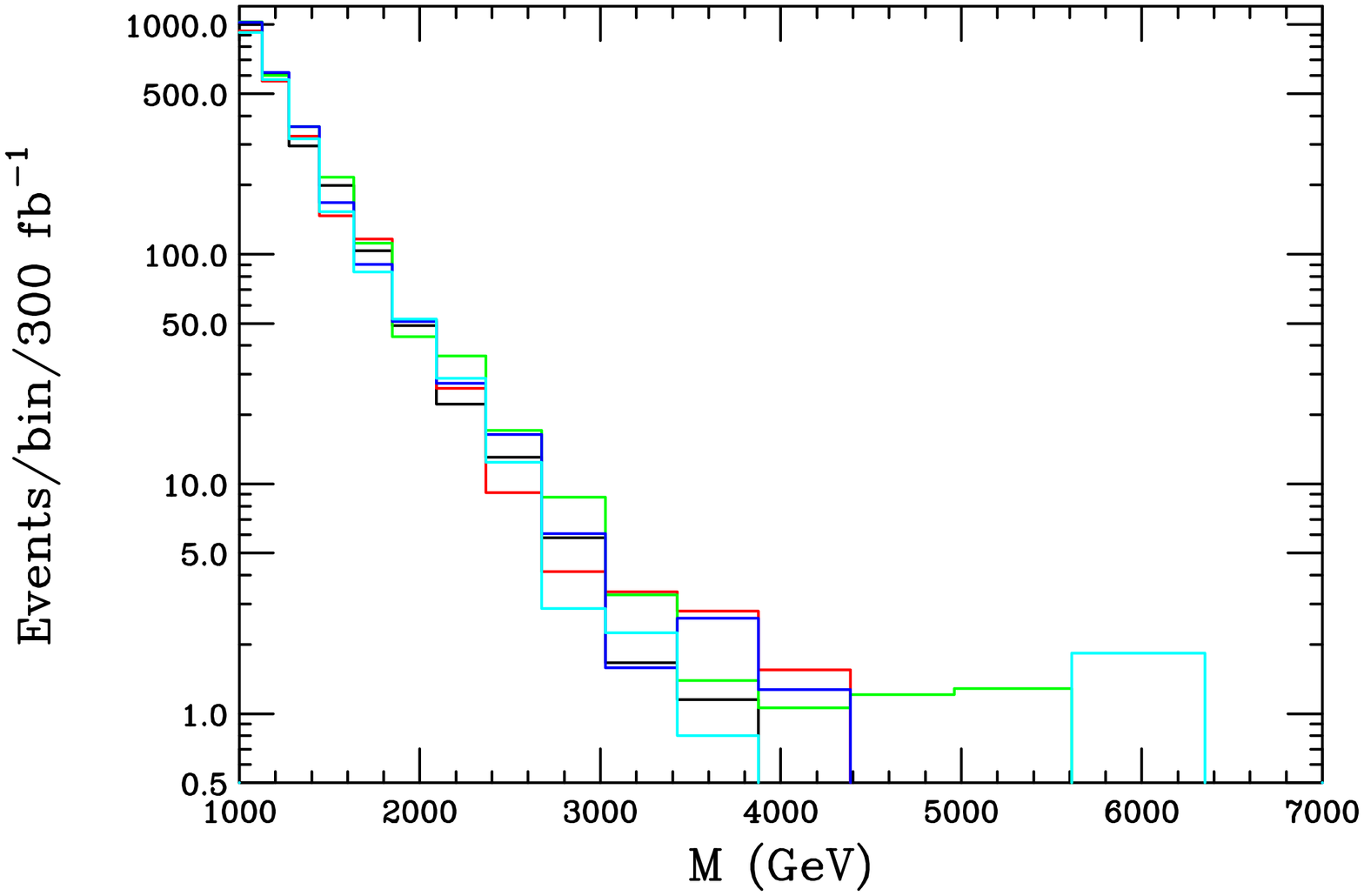}}
\caption{Same as the lower panel in previous figure but now showing the effect of increasing the width of all the bins by a factor of 5 to increase statistics.}
\label{fig25}
\end{figure}

In reality, however, the true experimental picture will be much more similar to the corresponding results shown in Fig.~\ref{fig2} where the statistical fluctuations 
are now explicitly included. Here we can still observe (at the $5\sigma$ level in comparison to the SM background) resonances 
themselves (at the corresponding mass reach values quoted above) but the 
statistical fluctuations 
in the data below the resonance peak now make all of the distributions look rather similar. A careful analysis does show some sensitivity to the higher mass 
resonances (which are quite close to being observable directly) in the case of KK gauge excitations as these have a very special associated signature due to the strong 
destructive interference in this mass range as discussed above{\cite {study}}. Even with the statistical fluctuations present 
it is possible to observe that the prediction for the gauge KK case lies systematically {\it below} that of the SM. 
One could argue that this bad situation with respect to the statistical fluctuations masking any significant effect might be improved by increasing the width 
of the dilepton mass bins to increase statistics and decrease the effects of fluctuations. 
In this case, however, we only gain in statistics at the price of reduced sensitivity as can be seen by the example shown 
in Fig.~\ref{fig25}, where the bin sizes are now increased by a relative factor of 5, with similar results found to hold in other rescaling cases. It thus seems quite 
unlikely that new resonances can be indirectly observed at the LHC except in exceptional circumstances where they have large effects on data in the dilepton mass 
range far below the resonance mass itself. 


\begin{figure}[htbp]
\centerline{
\includegraphics[width=8.5cm,angle=90]{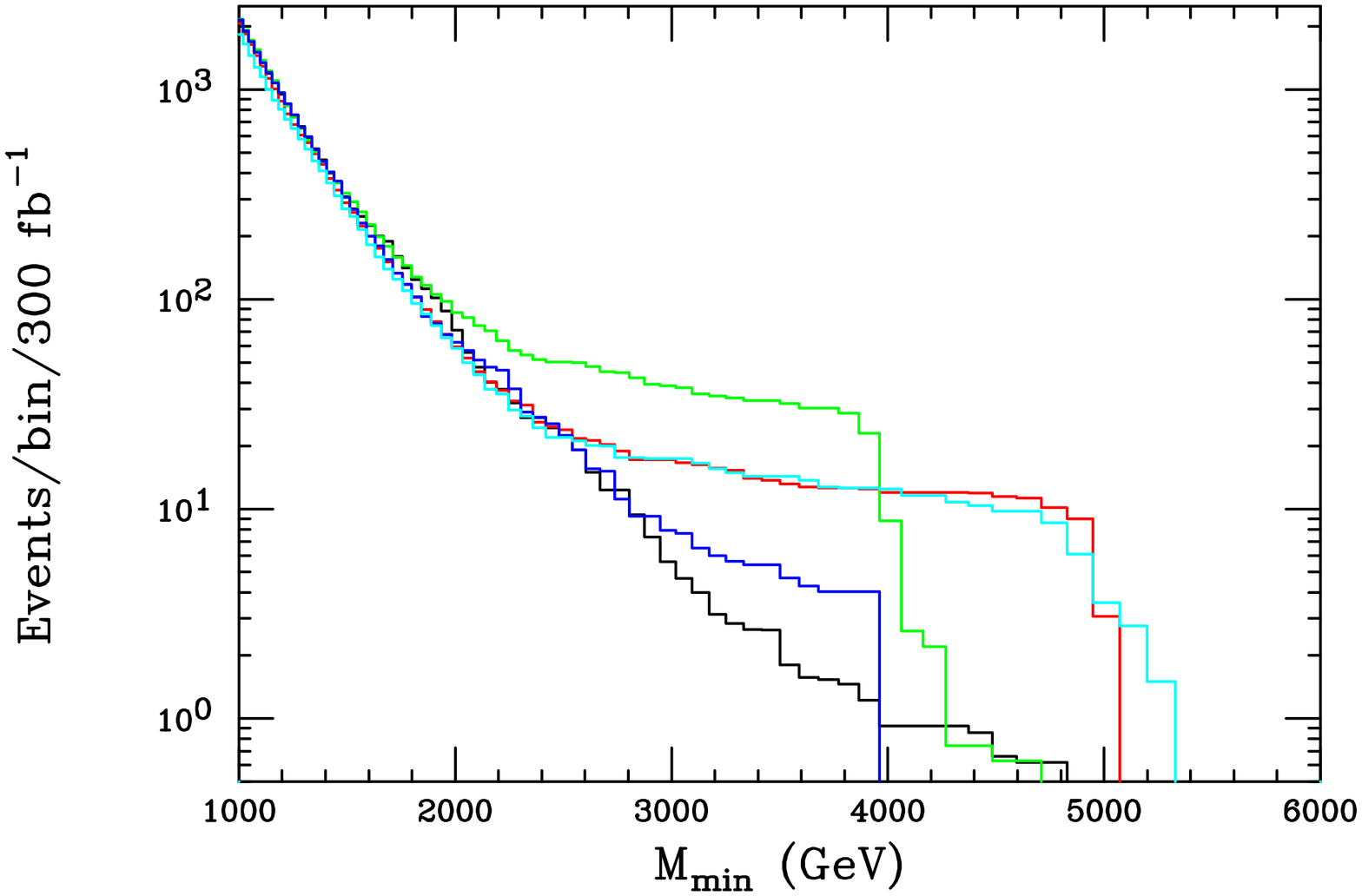}}
\vspace*{0.4cm}
\centerline{
\includegraphics[width=8.5cm,angle=90]{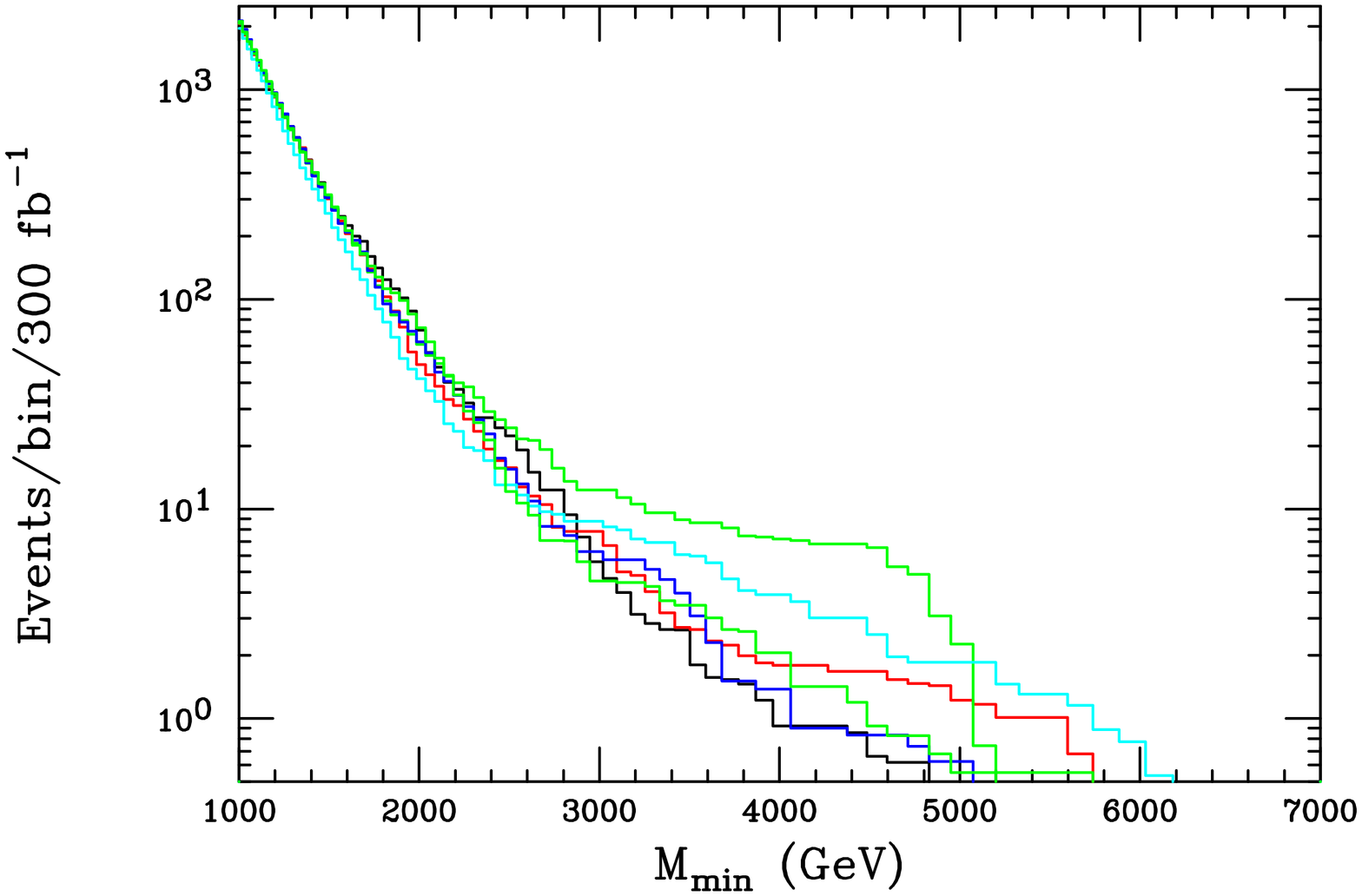}}
\caption{Same as in Fig.~\ref{fig2} but now showing the {\it cumulative} number of events above a minimum value of the dilepton invariant mass. In the lower panel an 
additional (lower) green histogram is also present for the $R$-parity violating sneutrino case assuming a resonance mass of 6 TeV. Note that statistical fluctuations 
are included here in both panels. }
\label{fig3}
\end{figure}

An alternative approach to using the $dN/dM$ distribution itself is to use the same data to form the associated cumulative distribution which displays the total number 
of events above some minimum dilepton mass cutoff. Such distributions certainly have added statistical power as well as increased sensitivity to the high mass region 
of interest to us and are not sensitive to bin-size effects. Examples of these types of distributions using our benchmark models for both sets 
of assumed masses are shown in Fig.~\ref{fig3}. In the top panel, 
all of the benchmark models are clearly seen above the SM background as would be expected from our earlier discussion. In the lower panel, the 5 TeV sneutrino and 
the 6 TeV gauge KK benchmarks may possibly lie significantly above the SM background while the other benchmarks appear to 
lie much lower. For comparison purposes the case of a 6 TeV sneutrino is also shown and the corresponding histogram is seen to lie quite close to the SM 
background expectations. In order to make a direct comparison of the benchmark models, consider, for example, the cumulative number of expected events we
find above a dilepton mass of 4 TeV (subject to very large statistical fluctuations as well as possibly significant systematic uncertainties) is 
1.0(1.8, 1.4, 3.9, 7.2, 1.7) for the SM(6 TeV SSM, 5 TeV RS graviton, 6 TeV gauge KK, 5 TeV sneutrino and 6 TeV sneutrino) cases, respectively, none of which remotely  
rises close to the level of a significant signal{\cite {fandc}} except possibly for the case of the 5 TeV sneutrino.  Clearly the cumulative distributions 
do not show any obvious evidence for sensitivity to new resonances beyond the direct LHC discovery reach even when potentially important 
systematic effects are neglected. This implies that except in exceptional circumstances new resonances cannot be indirectly detected using this method.

\begin{figure}[htbp]
\centerline{
\includegraphics[width=8.5cm,angle=90]{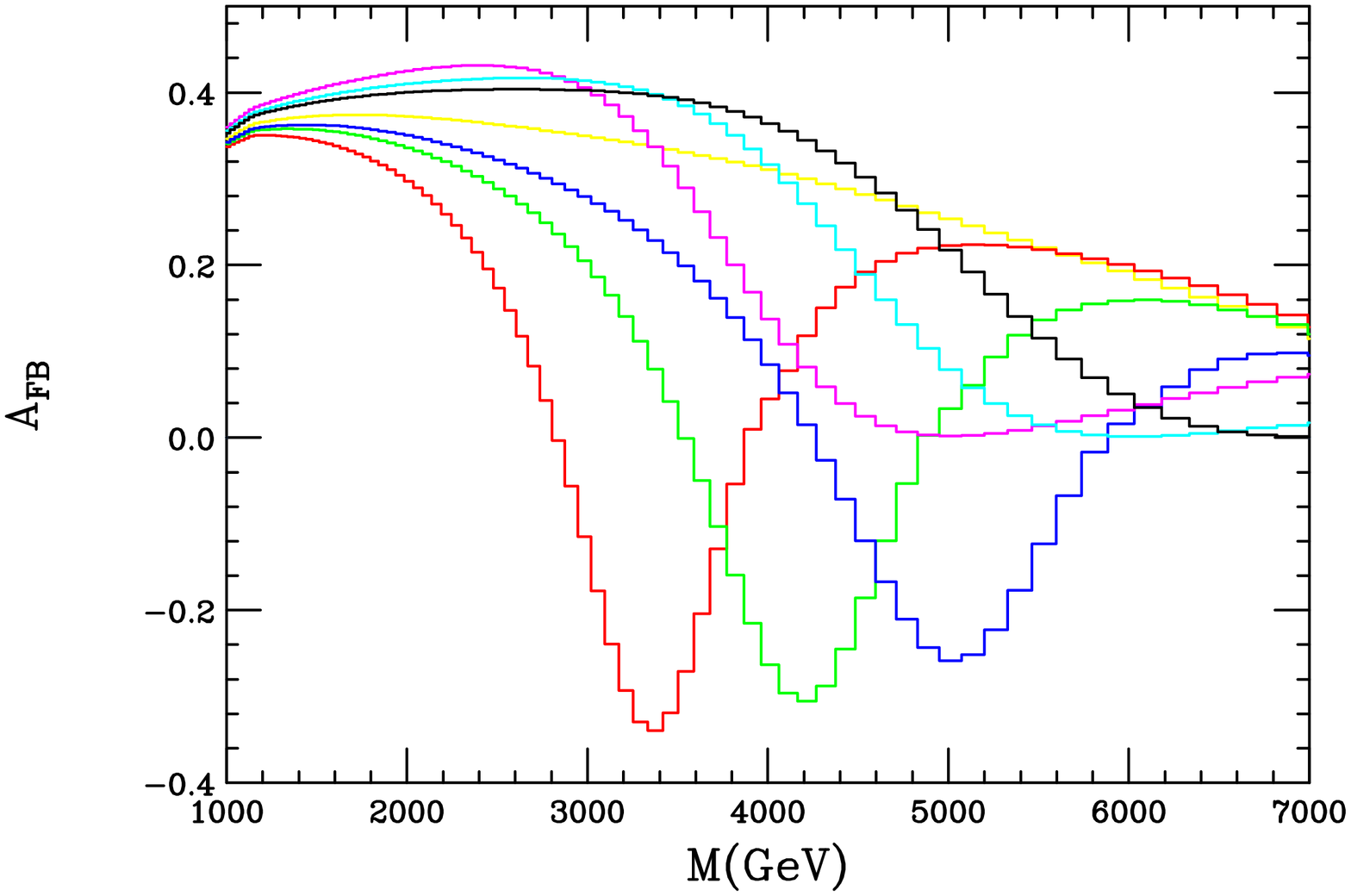}}
\vspace*{0.4cm}
\centerline{
\includegraphics[width=8.5cm,angle=90]{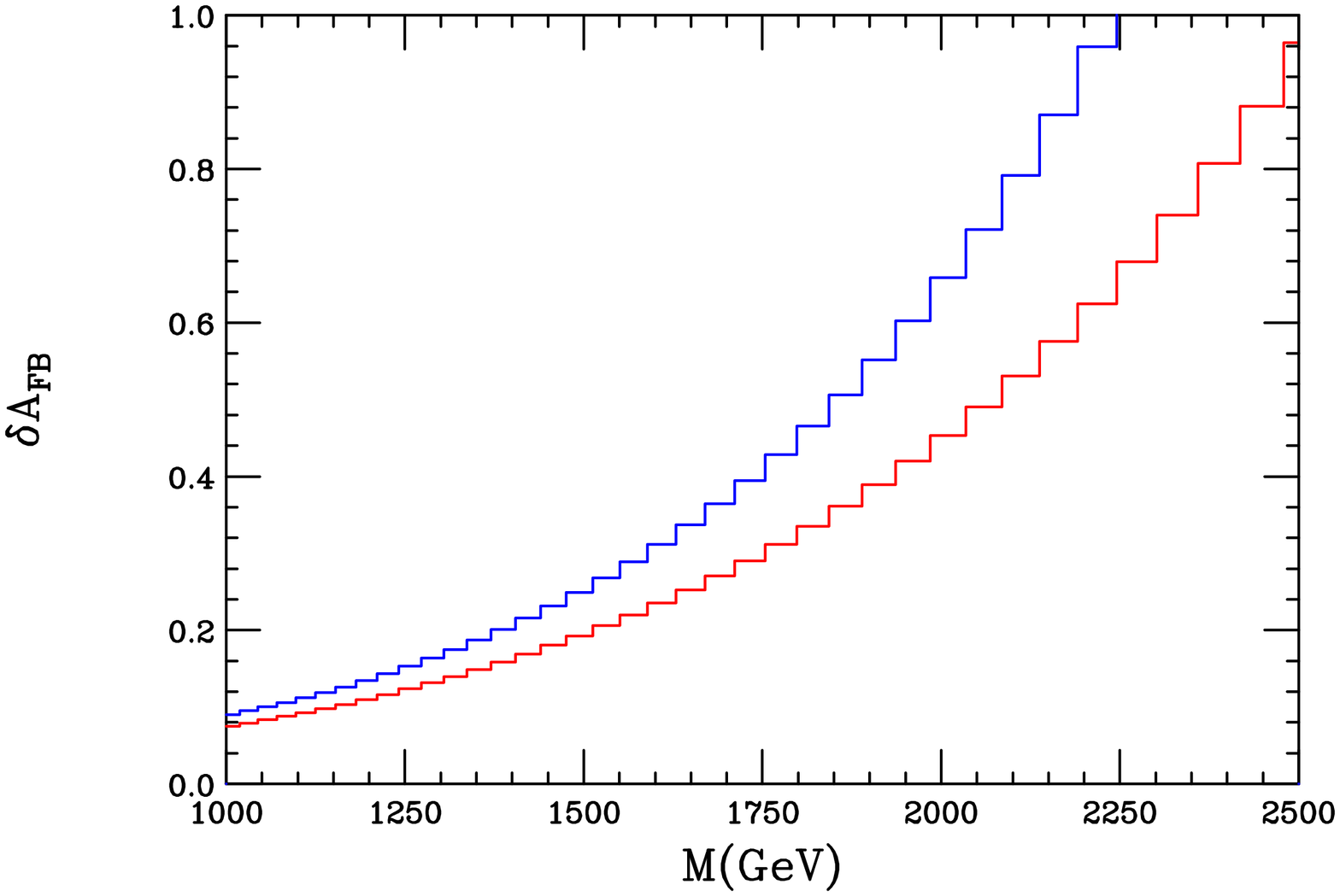}}
\caption{The upper panel shows the predicted values of $A_{FB}$ in the SM(yellow histogram), for the SSM with a $Z'$ mass of 4(5, 6) TeV corresponding to 
the red(green, blue) histograms, and for the case of gauge KK excitations with mass of 5(6, 7) TeV, corresponding to the magenta(cyan, solid) histograms, respectively. 
For purposes of demonstration, a cut of $|y| \geq 0.5$ has been applied in all cases. 
The lower panel shows the statistical error on the SM value of $A_{FB}$ as a function of the dilepton mass assuming a cut of $|y|\geq 0.5(0.8)$ 
corresponding to the red(blue) histogram. An integrated luminosity of 300 fb$^{-1}$ has been assumed in the lower panel.}
\label{fig4}
\end{figure}

Let us now return to the issue of the forward-backward asymmetry, $A_{FB}$, discussed above. In principle, this quantity shows a reasonable sensitivity to heavy 
resonances but is limited in applicability by reduced statistics. This can be most easily seen by examining the results presented in the top panel of Fig.~\ref{fig4} 
which shows (without errors) the expected values of $A_{FB}$ in the dilepton mass range of interest for the SM as well as for the SSM and gauge KK models for several 
different values of the resonance masses. Clearly $A_{FB}$ deviates significantly from the SM expectations in the region below the mass of the new resonance in all 
cases. However, as we will see, the statistical power of $A_{FB}$ in this same mass range is quite poor. 
To demonstrate the importance of a cut of $|y|\geq 0.8(0.5)$ on the available statistics we note that in the lowest dilepton mass bin 
considered here, near $\sim 1$ TeV,  we find 248 events in the SM case before applying this rapidity 
cut but only 105(155) events afterwards. In this mass bin, these cuts yield the asymmetry values of $A_{FB}=0.388\pm 0.090(0.346\pm 0.075)$ which show that 
the (statistical only) errors are already quite large. Above this dilepton mass the severity of the lower rapidity cut becomes far more significant and the size of 
the error on the asymmetry grows quite rapidly. {\footnote {It is important to remember that the severity of the rapidity cut increases with the dilepton mass 
due to the kinematic upper limit given in Eq. 1.}} 
This result is explicitly seen in the lower panel in Fig.~\ref{fig4} for either choice of the value of the lower cut 
on $|y|$. From this figure, however, we can easily imagine that if the mass 
of the new resonance were significantly less, $1.5-2$ TeV say, it would be possible that $A_{FB}$ measurements below the peak would provide significant sensitivity to 
new physics. In fact, it is well-known that for the new gauge boson case such data provides important information on the couplings of $Z'$-like states{\cite {rev}}.

By combining the $A_{FB}$ information for all dilepton masses above $\sim 1$ TeV we can determine the added sensitivity to new resonance states beyond the direct 
discovery search reach. Clearly such sensitivity will be dependent upon the specific nature of this new resonance and the precise lower bound on the rapidity applied 
to the final state leptons. However, scanning over the various benchmark models and possible resonance masses we find that the total contribution to the $\chi^2$ 
from the deviation in the value of $A_{FB}$ to be always less than $\sim 0.3-0.5\sigma$ and generally far smaller. This is not too surprising given the results shown in 
Fig.~\ref{fig4}. Thus it is quite clear that the additional information obtained 
from the dilepton angular distribution will not be very helpful in indirect searches for new resonances in the dilepton mass range of interest to us here above 4-5 TeV. 
Thus we conclude that such searches must essentially rely on the dilepton mass distribution except in very special circumstances.

\section{Discussion and Conclusions}

In this paper we have explored the possibility of indirectly discovering resonance states in the Drell-Yan channel at the LHC for particles of various spins when their 
masses are just beyond the direct discovery reach. Generally, we find this task to be quite difficult if not impossible even in the limit where systematic 
effects, such as PDF uncertainties, are neglected. The only exception to this result is the rather unusual circumstance of nearly degenerate photon and $Z$ 
Kaluza-Klein excitations. In particular, in the course of this analysis, we found that the forward-backward asymmetry, $A_{FB}$, of the dilepton pair was generally 
not a useful observable for performing such searches due to the large associated statistical errors in the mass range of interest above $\sim 2-2.5$ TeV. Essentially, 
these indirect searches must rely solely on the dilepton mass distribution. Except for possibly the 
special gauge KK case, where significantly modifications are observed in 
the dilepton invariant mass distribution at masses far below the peak due to strong interference with SM exchanges, new resonances are found to only perturb the 
mass distribution significantly in the neighborhood just below the peak itself where the statistics are commonly quite poor. 

Hopefully new Drell-Yan resonances will be relatively light and will soon be discovered at the LHC

\section{Acknowledgments}

The author would like to thank J. Wells for raising the important issue of indirect resonance searches at the LHC and for early discussions on the work presented here.
The author would also like to thank J.L. Hewett for discussions and the CERN Theory Group, where part of this work was done, for its hospitality. 

%
\def\MPL #1 #2 #3 {Mod. Phys. Lett. {\bf#1},\ #2 (#3)}
\def\NPB #1 #2 #3 {Nucl. Phys. {\bf#1},\ #2 (#3)}
\def\PLB #1 #2 #3 {Phys. Lett. {\bf#1},\ #2 (#3)}
\def\PR #1 #2 #3 {Phys. Rep. {\bf#1},\ #2 (#3)}
\def\PRD #1 #2 #3 {Phys. Rev. {\bf#1},\ #2 (#3)}
\def\PRL #1 #2 #3 {Phys. Rev. Lett. {\bf#1},\ #2 (#3)}
\def\RMP #1 #2 #3 {Rev. Mod. Phys. {\bf#1},\ #2 (#3)}
\def\NIM #1 #2 #3 {Nuc. Inst. Meth. {\bf#1},\ #2 (#3)}
\def\ZPC #1 #2 #3 {Z. Phys. {\bf#1},\ #2 (#3)}
\def\EJPC #1 #2 #3 {E. Phys. J. {\bf#1},\ #2 (#3)}
\def\IJMP #1 #2 #3 {Int. J. Mod. Phys. {\bf#1},\ #2 (#3)}
\def\JHEP #1 #2 #3 {J. High En. Phys. {\bf#1},\ #2 (#3)}

\end{document}